\ifx\pdfoutput\undefined        
  \documentclass[twocolumn]{article}
  \usepackage{url}
\else                           
  \documentclass[pdflatex,twocolumn]{article}
  \usepackage[bookmarks=false]{hyperref}
\fi

\usepackage{graphicx}
\usepackage{times}

\newcommand{\EQ}{\begin{equation}}
\newcommand{\EN}{\end{equation}}
\newcommand{\EQA}{\begin{eqnarray}}
\newcommand{\ENA}{\end{eqnarray}}

\newcommand{\Eq}[1]{Eq.~(\ref{#1})}
\newcommand{\Eqs}[2]{Eqs.~(\ref{#1}) and~(\ref{#2})}

\newcommand{\eqs}[2]{(\ref{#1}) and~(\ref{#2})}

\newcommand{\Fig}[1]{Fig.~\ref{#1}}

\newcommand{\dd}{{\rm d} {}}

\newcommand{\ypnas}[5]{ (#1) #5, {\em Proc.\ Natl.\ Acad.\ Sci.\ }{\bf #2}, #3--#4.}
\newcommand{\yjacs}[5]{ (#1) #5, {\em J.\ Am.\ Chem.\ Soc.\ }{\bf #2}, #3--#4.}

\newcommand{\ysci}[5]{ (#1) #5, {\em Science }{\bf #2}, #3--#4.}

\newcommand{\ynat}[5]{ (#1) #5, {\em Nature }{\bf #2}, #3--#4.}

\newcommand{\yoleb}[5]{ (#1) #5, {\em Orig.\ Life Evol.\ Biosph.\ }{\bf #2}, #3--#4.}

\newcommand{\yjour}[6]{ (#1) #6, {\em #2} {\bf #3}, #4--#5.}
\newcommand{\yjourS}[6]{ (#1) #6 {\em #2} {\bf #3}, #4--#5.}

\newcommand{\ybook}[3]{ (#1) {\em #2}, #3.}

\newcommand{\poleb}[2]{ (#1) #2, {\em Orig.\ Life Evol.\ Biosph.} (in press).}

\newcommand{\ea}{{\em et al., }}
\newcommand{\eaa}{{\em et al. }}
\def\half{{\textstyle{1\over2}}}

\begin{document}

\title{Uni-directional polymerization leading to homochirality
in the RNA world}
\author{M.\ Nilsson$^{1,2}$, A.\ Brandenburg$^1$, A.\,C.\ Andersen$^1$ and
S.\ H\"ofner$^3$}
\date{\small
$^1$Nordita, Blegdamsvej 17, DK-2100 Copenhagen \O, Denmark\\
$^2$Department of Physical Resource Theory, Chalmers University of Technology, SE-412 96 G\"oteborg, Sweden \\
$^3$Department of Astronomy \& Space Physics, Uppsala University, Box 515, SE-751 20 Uppsala, Sweden}

\maketitle

\abstract{
The differences between uni-directional and bi-directional polymerization
are considered.
The uni-directional case is discussed in the framework of the RNA world.
Similar to earlier models of this type, where polymerization was assumed
to proceed in a bi-directional fashion (presumed to be relevant to peptide
nucleic acids), left-handed and right-handed monomers are produced via
an autocatalysis from an achiral substrate.
The details of the bifurcation from a racemic solution to a
homochiral state of either handedness is shown to be remarkably independent
of whether the polymerization in uni-directional or bi-directional.
Slightly larger differences are seen when dissociation is allowed and
the dissociation fragments are being recycled into the achiral substrate.
}

\section*{Introduction}

The origin of homochirality is usually believed to be closely
connected with the origin of life (see Bada 1995 for an overview).
It may have even been a {\it prerequisite} for life in that the structural
stability provided by chiral polymers may have been essential for the
assembly of the first replicating molecule.
If this is so, it would probably mean that the origin of homochirality
had to be a physical one.
Possible candidates for a physical origin of homochirality include the
presence of polarized light from a nearby neutron star
(Rubenstein et al.\ 1983),
magnetic fields (Thiemann 1984, Rikken and Raupach 2000), or mechanisms
involving the electroweak force (e.g., Hegstrom, 1984).
However, Bailey et al.\ (1998) and Bailey (2001) showed later that
supernova remnants have not actually displayed circularly polarized light.
Another perhaps more likely possibility is that homochirality developed
rather as a {\it consequence} of life.
This would mean that some primitive form of life should have been possible
without chirality having played any role in this.

In connection with the origin of life one used to discuss the hypothesis
of a relatively simple self-replicating molecule (e.g.\ Frank 1953).
This picture ignores the possible importance of compartmentalization
that may be required for achieving the concentrations necessary for the
chemical reactions to take place.
This led to the concept of a very early lipid world that would have
preceded the often discussed RNA world.
Some insight into these ideas can be gained by looking at
recent theoretical attempts to build life from scratch invoking a series of steps
and chemical processes that are thermodynamically possible
(Rasmussen \ea 2003).
Interestingly enough, their approach involves peptide nucleic acid
(PNA) because of its charge carrying properties and the molecule's hydrophobic backbone.
Its potential as contemporary genome, which would for example require a machinery for protein transcriptase,  was not utilized at this stage, although
it may undoubtedly become a candidate for carrying genetic information at
later evolutionary stages.

Although this is speculation and details are unknown, the idea of a
combined PNA/lipid world provides
an attractive scenario for discussing the origin of homochirality in the
context of genetic evolution (Nelson \ea 2000, Pooga \ea 2001).
We picture here a situation where PNA has developed to having autocatalytic
properties, just like RNA in the RNA world (Woese 1967).
PNA can be achiral if its peptide backbone is derived from glycine.
The step toward a chiral backbone invoking, for example, poly-alanine,
seems like a relatively minor one (one of two H is being substituted
by CH$_3$ in the CH$_2$ piece).
However, there are two different ways of doing this, leading to either
{\sc l} or {\sc d} alanine.
The assembly of mixed {\sc l} and {\sc d} PNA poly-nucleotides is unlikely,
just as it is unlikely in the corresponding case of DNA polymerization
(Joyce \ea 1984).
Moreover, the addition of a nucleotide of opposite handedness is known
to `spoil' further polymerization (also known as `enantiomeric
cross-inhibition').
This makes it increasingly unlikely to generate {\sc l} and {\sc d} polymers of
any appreciable length greater than just a few.

The main difference between PNA and DNA polymerization is that DNA
can attach new monomers only on the 3' end of the ribose sugar
(e.g.\ Turner \ea 2000), so polymerization is uni-directional and can
only proceed in one direction.
By contrast, PNA does not have this restriction and can polymerize
in a bi-directional fashion in either direction.
The latter case has been addressed in a number of recent studies
starting with Sandars (2003), but the former case is more
readily amenable to laboratory verification, as is shown by recent
experiments confirming the process of enantiomeric cross-inhibition
(Schmidt et al.\ (1997) and Kozlov et al.\ 1998).
Given that the differences between uni-directional and bi-directional
polymerization have not yet been explored, we must first extend the
formalism of Sandars (2003) to the uni-directional case and focus then
on the comparison between the two.

Although enantiomeric cross-inhibition seems to be an important
ingredient of homochirality, this can only work if the production of
new monomers is somehow biased toward the enantiomeric excess of the
already existing polymers -- even if this bias is extremely tiny.
This is the second important ingredient of homochirality and is
known as {\it autocatalysis}.
Certain chemical reactions are indeed known to have such properties
(Soai \ea 1995, Sato et al.\ 2003, Mathew \ea 2004).
It is important to point out that these reactions are only based on
dimerization, but they are nevertheless quite valuable in establishing
the basic elements of homochiralization in chemical systems
(Kitamura et al.\ 1998, Plasson et al.\ 2004), and can lead to
quantitative predictions (Blackmond et al.\ 2001, Buono and Blackmond 2003).
For a recent review see the paper by Mislow (2003).

The importance of the combined action of enantiomeric cross-inhibition
on the one hand and autocatalysis on the other has been well known
since the very early work of Strong (1898) and a
seminal paper of Frank (1953), who first proposed a
simple mathematical model consisting of only two variables representing
the relative numbers of left and right handed building blocks.
His paper was tremendously insightful in that he understood not only the
two basic ingredients needed for homochirality, but he was also aware
that there are two rather different scenarios through which homochirality
can be achieved, depending basically on how frequent the creation of a
potential life bearing molecule is.
If the creation of life was sufficiently frequent, life may have emerged
at different locations on the Earth's surface (including the oceans),
giving rise to the interesting possibility of having different life
forms of opposite handedness simultaneously.
This is the case studied recently by Brandenburg and Multam\"aki (2004),
who estimated that left and right handed life forms could have coexisted
for not more than the first 500 Million years.
This is because different populations will spread over the Earth's surface
and come eventually into contact, extinguishing one of the two life forms.
The other possibility is that the creation of life was an infrequent
event, in which case there was ever only one life form, which was then the
one that led eventually to the global population over the Earth's surface.
Regardless of which of the two scenarios applies, the final outcome would
have been the same.

In his paper, Frank (1953) only analyzed the second alternative in detail.
This is also the scenario discussed in most of the approaches since then,
which all discuss homochirality as the result of a bifurcation process
[see also Saito and Hyuga (2004a) for a recent classification of different
possibilities].
This forms also the basis for the model discussed in the present paper,
where we present a modification of a detailed polymerization model
proposed recently by Sandars (2003).
In this model the enantiomeric excess grows exponentially in time.
However, if the creation of life is a frequent event, the process
toward global homochirality can only occur linearly in time
(Brandenburg and Multam\"aki 2004; see also Saito and Hyuga 2004b for
related work).

In the model by Sandars (2003),
autocatalysis is incorporated by assuming that the rate of monomer
production of given handedness is proportional to the concentration of
polymers of the same handedness.
As noted above, this effect alone, i.e.\ without the additional effect
of enantiomeric cross-inhibition, cannot lead to complete homochirality,
because the initial enantiomeric excess is not (or only weakly) amplified.
In order to model this quantitatively, Sandars (2003) assumed that
polymerization can, at a certain rate, also occur with monomers of
opposite handedness.
This reaction produces chemically inactive products and it acts thus
as a means of removing oppositely oriented building blocks (that are
already in the minority) from the system.
This model has been studied further by Wattis and Coveney (2005) and
by Brandenburg \eaa (2005a, hereafter referred to as BAHN) who showed
that, for large enough fidelity of the catalyst, the departure from a
homochiral state occurs exponentially fast at a growth rate that depends
on the fidelity and the rate of enantiomeric cross-inhibition.
They also discussed a model consisting only of primers and dimers
which can be reduced to a set of two ordinary differential equations
which are similar to those of Frank (1953).
An important difference to Frank's model is the form of the
cross-inhibition term.
As discussed by Blackmond (2004), the feedback term in his model
corresponds to the formation of inactive dimers with one left and one
right handed building block. This is unrealistic, because dimers with
two left or two right handed building blocks should also form.
This let her to the conclusion that the dimers must act as catalysts.

We have emphasized that
the original model of Sandars assumed that polymerization can occur on
either end of the polymer.
While this may be a reasonable assumption in general (and probably also
for PNA), it is not realistic for RNA polymerization where polymerization
can usually only proceed in a uni-directional fashion.
Since uni-directional polymerization leads to a simpler model, and since
the derivation of the bi-directional polymerization model has already
been discussed elsewhere (see, e.g., BAHN), the uni-directional case
is ideal for introducing the basic ingredients of the model.
Following the mathematical description of the uni-directional model, we
present numerical solutions that show that the main conclusions obtained
from the earlier bi-directional polymerization models carry over to the
uni-directional case.
This addresses possible objections that the Sandars model is not
applicable to RNA and DNA polymerization that is more easily
amenable to detailed laboratory verification.

\section*{Polymerization model}
\label{polymerization}

The starting point of the model is a basic polymerization process

\begin{equation}
	L _{n} + L_{1} \stackrel{ k_{S}~}{\longrightarrow}  L_{n+1} ,
\end{equation}
where $L_{n}$ denotes left handed polymers of length $n$ and $k_{S}$ the reaction rate. The corresponding model of the polymerization process reads

\begin{equation}
	{\dd\over\dd t}[L_n] = -k_{S} [L_{1}] \left( [L_{n}] - L_{n-1}\right),
\label{Lneqn}
\end{equation}
where $[ L_{n} ]$ is the concentration of $L_{n}$.
New building blocks are continuously 
added to the model, e.g.\ by the inclusion of a substrate
that provides a source $Q_L$ of new monomers, i.e.\
\begin{equation}
	{\dd\over\dd t}[L_1]  = Q_{L} - \sum _{n=1}^N k_{S} [L_{1}] [L_{n}] 
\label{L1eqn}
\end{equation}

\begin{figure}[t!]\begin{center}
\includegraphics[width=\columnwidth]{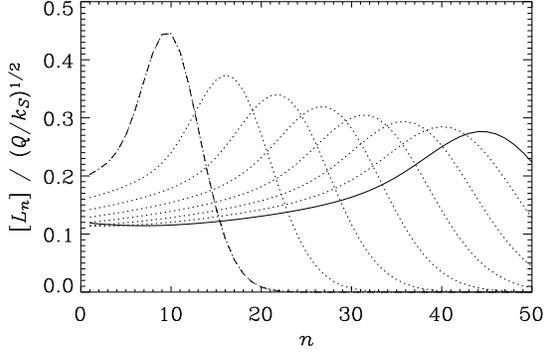}
\end{center}\caption[]{
Wave-like propagation of a finite amplitude perturbation
in the uni-directional polymerization model.
The initial profile is a gaussian.
Note the undisturbed outward propagation of the wave at $n=N$.
The time difference between the different curves is $20/(k_SQ)^{1/2}$.
We have shown the first and last times as dashed and solid lines,
respectively, and all other times as dotted lines.
The parameters are $N=50$ and $k_C/k_S=1$.
}\label{pwave}\end{figure}

The solution of Eqs.~(\ref{Lneqn}) and (\ref{L1eqn})
is simply a wave traveling toward longer polymers at velocity $k_{S}[L_1]$
(see \Fig{pwave}), as can also be seen by considering the continuous limit
of this equation,
$\partial[L_n]/\partial t=-k_{S}[L_1]\partial[L_n]/\partial n$.
Note that, in contrast to a similar result for bi-directional
polymerization (see BAHN, their Fig.~1), the functional form of $[L_n]$
is continuous between $n=1$ and $n=2$.
In the bi-directional case $[L_1]$ is about twice as large as $[L_2]$.

The model becomes more interesting when the right handed polymers, $R_n$,
are also included. The interaction between the mirrored strands is assumed
to occur through two separate phenomena: enantiomeric 
cross-inhibition and enzymatic autocatalysis. The autocatalysis makes the left handed, respective right handed, polymers catalyze the production of left, respective right, handed building blocks. 
The source terms $Q_L$ and $Q_R$ are proportional to the concentration of the
achiral substrate $[S]$ and a corresponding reaction coefficient $k_{C}$.
In the case of perfect fidelity, $f=1$, the source terms are written as
\begin{equation}
	Q_L=k_C[S]C_L,\quad
	Q_R=k_C[S]C_R,
\end{equation}
where $C_L$ and $C_R$ are some measures of the catalytic effect of the
already existing left-handed and right-handed polymers.
The should be a monotonically increasing function of the
overall concentration of the left handed polymers. 
The exact functional form of these expressions are not known.
In fact, different authors have chosen different prescriptions for
$C_L$ and $C_R$.
The qualitative results of the models do however not seem not affected by this choice.
We find it natural to assume that
\begin{equation}
	C_{L} = \sum _{n=1}^{N} n [ L_{n} ]  ,
\end{equation}
\begin{equation}
	C_{R} = \sum _{n=1}^{N} n [ R_{n} ]   .
\end{equation}

In the more general case of finite fidelity of the assumed autocatalysis,
i.e.\ for $0<f<$, we model there will be `cross-talk' between between the
two handednesses, so we write
\begin{equation}
	Q_L=k_C[S]\Big\{\half(1+f)C_L+\half(1-f)C_R+C_{0L}\Big\},
	\label{QLdef}
\end{equation}
\begin{equation}
	Q_R=k_C[S]\Big\{\half(1+f)C_R+\half(1-f)C_L+C_{0R}\Big\} , 
\label{QRdef}
\end{equation}
Here the terms $C_{0L}$ and $C_{0R}$ allow for the possibility of
non-catalytic production of left and right handed monomers.
However, in the following we assume $C_{0L} = C_{0R} =0$.
(The inclusion of $C_{0L}$ and $C_{0R}$ terms leads to so-called imperfect
bifurcations; see Fig.~6 of BAHN.)

The enantiomeric cross-inhibition occurs when a building block attaches to a polymer of the
opposite handedness. The resulting polymer cannot continue to grow at the affected end and can therefore be considered spoiled. This phenomenon has been observed in experiments by
(Joyce \ea 1984) who studied template-directed polymerization. When the cross-inhibition is
included, the set of reactions in the model is (for $n\ge2$)
\begin{eqnarray}
L_{n}+L_1&\stackrel{k_S~}{\longrightarrow}&L_{n+1},
\label{react1}\\
L_{n}+R_1&\stackrel{k_I~}{\longrightarrow}&L_nR_1,
\label{react2}
\end{eqnarray}
and for all four equations we have the complementary reactions
obtained by exchanging $L$ and $R$. The new parameter $k_{I}$ measures
the rate at which the cross-inhibition occurs.
The rate equations now read (for $n\ge2$)
\begin{equation}
{\dd[L_{n}]\over\dd t}=k_S[L_1]\left([L_{n-1}]-[L_{n}]\right)
-k_I[L_{n}][R_1],
\label{Ln_new2}
\end{equation}
\begin{equation}
{\dd[R_{n}]\over\dd t}=k_S[R_1]\left([R_{n-1}]-[R_{n}]\right)
-k_I[R_{n}][L_1].
\label{Rn_new2}
\end{equation}
The evolution of the spoiled polymers, $L_{n}R_{1}$ and $R_{n}L_{1}$,
can be discarded, because, in contrast to bi-directional polymerization,
their concentrations do not enter the uni-directional model.

In comparison with bi-directional polymerization we note that here
for $n=2$ there is no extra 1/2 factor in front of the $[L_1]^2$ and
$[R_1]^2$ terms in \Eqs{Ln_new2}{Rn_new2}.
This is because with polymerization from either end the total reaction
rate would be twice as big.
However, when two monomers interact, the corresponding reaction equation
is the same for uni-directional and bi-directional polymerization, because
the two reacting monomers are indistinguishable.
Thus, whether the first binds to the second or the second to the first
monomer does not make a difference.
This is then equivalent to saying that for two monomers polymerization
can occur both on the 3' and on the 5' end of the ribose sugar.
In effect, this removes an awkward 1/2 factor for the $n=2$ equations
in the model of Sandars (2003); see also Eq.~(7) of BAHN.

The reactions \eqs{react1}{react2} imply the presence of additional loss
terms in the evolution equations of monomers, so instead of \Eq{L1eqn}
we now have
\begin{equation}
	{\dd\over\dd t}[L_1]=Q_{L}-\lambda_L[L_1],
\label{L1eqn2}
\end{equation}
\begin{equation}
	{\dd\over\dd t}[R_1]=Q_{R}-\lambda_R[R_1],
\label{R1eqn2}
\end{equation}
where we have defined decay rates
\begin{equation}
\lambda_L=k_{S}\left([L_1]+\sum_{n=1}^N [L_{n}]\right)
+k_{I}\sum_{n=1}^N [R_{n}],
\label{lamL}
\end{equation}
\begin{equation}
\lambda_R=k_{S}\left([R_1]+\sum_{n=1}^N [R_{n}]\right)
+k_{I}\sum_{n=1}^N [I_{n}].
\label{lamR}
\end{equation}
Comparing again with the bi-directional model, the present model has an
extra $[L_1]$ (or $[R_1]$) term, but there is no factor of 2 in front
of the $k_S$ and $k_I$ terms and the sums over the concentrations of
semi-spoiled polymers are also absent.

From symmetry considerations it follows that there always exist a racemic
steady state  ($[R_n]=[L_n]$) of the rate equations. In fact, we can show that a steady 
state is given by (for $n\ge2$)
\begin{equation}
[L_n]= \left( 1+ \frac{k_I}{k_S} \right) ^{-(n-1)}[L_1]
\quad\mbox{(racemic)},
\label{racemic}
\end{equation}
In particular, if $k_I=k_S$, then $[L_n]=[L_1]/2^{n-1}$,
i.e.\ $[L_n]$ drops by a factor of 2 from one $n$ to the next.
This is also true between $[L_1]$ and $[R_1]$, while in the bi-directional
model their ratio is 4.

\begin{figure}[t!]\begin{center}
\includegraphics[width=\columnwidth]{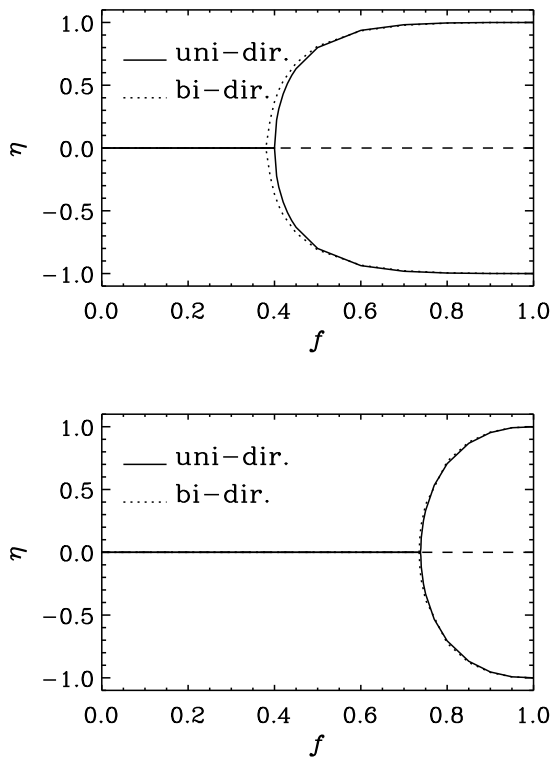}
\end{center}\caption[]{
Bifurcation diagram for two different values of $k_I/k_S$ (=1 in the upper
panel and 0.1 in the lower panel).
Note the transition from a racemic to homochiral state as a function of 
the autocatalytic fidelity $f$. Homochirality is measured in terms of $\eta =
(\sum _{n} n ([ L_{n} ] - [ R_{n} ])/(\sum _{n} n ([ L_{n} ] + [ R_{n} ])$.
For weak enantiomeric cross-inhibition ($k_I/k_S=0.1$ in the lower panel)
the range of permissible values of the fidelity parameter is decreased,
demonstrating the importance of enantiomeric cross-inhibition.
}\label{bifurc}\end{figure}

While the existence of a racemic solution is trivial, the interesting question is whether there
exist other fixed points of the equations, and in this case which of these fixed points are stable
under certain conditions. As was shown in BAHN the model typically 
goes through a pitchfork bifurcation from a single stable fixed point (the racemic solution) to
a state with two homochiral stable fixed points where the racemic solution corresponds to an unstable fixed point.
The order parameter controlling the bifurcation is the fidelity $f$
of the autocatalysis.
In \Fig{bifurc} we show the enantiomeric excess, defined here as
\begin{equation}
\eta={C_R-C_L\over C_R+C_L},
\label{etdef}
\end{equation}
for $k_I/k_S=1$ and $k_I/k_S=0.1$.
We also compare with the corresponding result from the bi-directional
polymerization model.
The difference between the two cases is however surprisingly small.

\section*{Polymer dissociation}

The model described in the last section provides a possible explanation of homochirality, without
appealing to external mechanisms for the symmetry breaking. One may also argue that the model
is rather realistic in that it explicitly considers the polymerization process. Less satisfactory are some 
of the details in the description of the polymerization process. Perhaps most importantly, the polymerization process is
irreversible, no chain-breaking is included in the model.
As we have already pointed out in an earlier paper
(Brandenburg \ea 2005b, hereafter referred to as BAN), this is unrealistic
because for large enough fidelity the polymer length always tends to diverge.
Also, the model cannot be
self-contained since there is no feedback from the polymers back to the substrate.

Before discussing in more detail the differences between uni-directional and
bi-directional polymerization in the presence of dissociation, let us first
recall the main aspects of the polymerization model with dissociation, as
developed recently by BAN.
The dissociation process is described by the reaction
\begin{eqnarray*}
L_n &\stackrel{\gamma_S~}{\longrightarrow} & L_m + L_{n-m},
\end{eqnarray*}
and the corresponding reaction for the right handed polymers. It turns out that there are a 
number of subtleties that need consideration when constructing the detailed model of the 
chain breaking. For example, if we assume that the fragments can continue to 
polymerize, the result is a catastrophic over-abundance of the short chains. The reason for this is
that all building blocks ($L_{1}$ and $R_{1}$) are used to produce longer polymers whereas 
polymers of length two or more cannot (according to the reactions above) agglomerate into longer polymers. One way of remedy would of course be to include the agglomeration in the model, 
but the disadvantage of this is that the model then becomes significantly more complex due to the
higher degree of nonlinearity.
These issues are discussed in further detail in BAN, where also
a number of possible of the model model are considered.
We focus here on the model where the polymerization fragments
are recycled back into the achiral substrate.
In the rest of this paper we discuss the modifications necessary to
incorporate dissociation in a uni-directional polymerization model.

\begin{figure}[t!]\begin{center}
\includegraphics[width=\columnwidth]{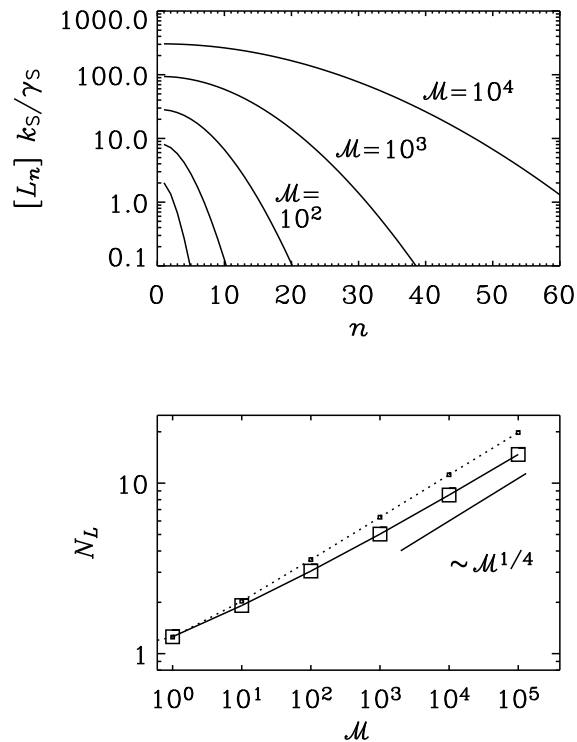}
\end{center}\caption[]{
Isotactic equilibrium states with polymerization, dissociation, and
recycling of fragments into the substrate, for different values of
${\cal M}$ (upper panel), and the mean polymer length $N_L$ (lower panel,
solid line), compared with the bi-directional polymerization model of
BAHN (dotted line).
}\label{pnl}\end{figure}

In the presence of dissociation, the new system of equations is
\begin{eqnarray*}
{\dd\over\dd t}[L_n] & = & p_n^{(L)}-(n-1)\gamma_S[L_n], \\
{\dd\over\dd t}[R_n] & = & p_n^{(R)}-(n-1)\gamma_S[R_n],
\end{eqnarray*}
where $p_n^{(L)}$ and $p_n^{(R)}$
indicate the terms due to polymerization described above. The source term in the substrate
is given by
\begin{equation}
Q=W_L+W_R+W_{LR}+W_{RL}
\end{equation}
where
\begin{equation}
W_L=\sum_{n=1}^N n w_n^{(L)},\quad
W_R=\sum_{n=1}^N n w_n^{(R)},
\end{equation}
is the total number of recycled building blocks (both left-handed and
right-handed), and
\begin{equation}
W_{LR}=\sum_{n=1}^N (n+1) w_n^{(LR)},\quad
W_{RL}=\sum_{n=1}^N (n+1) w_n^{(RL)}
\end{equation}
are the corresponding contributions from fragmented (inactive) polymers.

Like in the bi-directional case, the
average polymer length scales with a quarter power of the parameter ${\cal M}=
(k_{S} / \gamma _{S})\sum _{n=1}^N n \left( [L_{n}] + [ R_{n}] \right)$.
Thus, in order to achieve appreciable polymer length, the
normalized total mass must be sufficiently large.

Histograms of the chain distribution and the dependence of the chain
length on the total normalized mass are given in \Fig{pnl} and compared
with the bi-directional case.
For small chain mass (${\cal M}\leq10$) the chains tend to be very
short ($N_L\approx1...2$), which is common to both bi-directional and
uni-directional cases.
For larger total mass, however, the two cases begin to depart from each
other such that for the same total mass the chains are slightly shorter
in the uni-directional case.

\section*{Conclusions}

In the present paper we have modified the polymerization model of Sandars
(2003) such that polymerization is only possible on one of the two ends
of the polymer.
Although PNA polymerization is probably still bi-directional,
this is normally not the case for RNA polymerization.
The significance of considering RNA polymerization is that it is readily
amenable to direct experimental verification (e.g., Joyce 1984).
One of the perhaps most curious properties of the model is the
wave-like evolution of the polymer length after initializing the
polymerization process.
This prediction could possibly be tested experimentally by setting
up a range of different polymerization experiments that are being
stopped at different times.
A subsequent analysis, as it is done for DNA sequencing, might then reveal
a structure as seen in \Fig{pwave}.

We emphasize that homochirality appears spontaneously when two separate mechanisms are present in the polymerization process: autocatalysis and enantiomeric cross-inhibition. The accuracy of the
autocatalysis is parameterized by a fidelity factor. At low fidelity the polymerization leads to
a racemic solution whereas at higher fidelity a homochiral state is reached from an initially (almost) racemic solution. The corresponding bifurcation diagram displays a classic pitchfork bifurcation
and the autocatalytic fidelity acts as a control parameter.
The differences between uni-directional and bi-directional polymerization
are however surprisingly small.

In the second part of this paper we have extended the model to include
dissociation within the framework of uni-directional polymerization.
As in the case of bi-directional polymerization,
the model becomes chemically more realistic in that longer chains are
now possible.
Moreover, the model is constructed to be self-contained
in that the need for external
replenishing of the substrate is now replaced by the recycling
of dissociation fragments.
With respect to chirality, the qualitative behavior of the model
is shown to persist the inclusion of dissociation. We therefore conclude
that the existence of a transition between a racemic and homochiral state, as a function of the 
autocatalytic fidelity, is a robust phenomenon within the class of models under  consideration.

\section*{References}

\begin{list}{}{\leftmargin 3em \itemindent -3em\listparindent \itemindent
\itemsep 0pt \parsep 1pt}\item[]

Bada, J. L.\ynat{1995}{374}{594}
{595}{Origins of homochirality}

Bailey, J., Chrysostomou, A., Hough, J. H., Gledhill, T. M., McCall,
A., Clark, S., M\'enard, F. and Tamura, M.\ysci{1998}{281}{672}
{674}{Circular polarization in star forming regions: implications for
biomolecular homochirality}

Bailey, J.\yoleb{2001}{31}{167}
{183}{Astronomical sources of circularly polarized light and the origin
of homochirality}

Blackmond, D. G.\ypnas{2004}{101}{5732}
{5736}{Asymmetric autocatalysis and its implications for the origin
of homochirality}

Blackmond, D. G., McMillan, C. R., Ramdeehul, S., Schorm, A., and
Brown J. M.\yjacs{2001}{123}{10103}
{10104}{Origins of asymmetric amplification in autocatalytic alkylzinc 
additions}

Brandenburg, A., Andersen, A., H\"ofner, S., and Nilsson, M.\poleb{2005a}
{Homochiral growth through enantiomeric cross-inhibition}
Preprints available online at: \url{http://arXiv.org/abs/q-bio/0401036} (BAHN).

Brandenburg, A., Andersen, A., and Nilsson, M.\poleb{2005b}
{Dissociation in a polymerization model of homochirality}
Preprints available online at: \url{http://arXiv.org/abs/q-bio/0502008} (BAN).

Brandenburg, A., \& Multam\"aki, T.\yjourS{2004}{Int.\ J.\ Astrobiol.}{3}{209}
{219}{How long can left and right handed life forms coexist?}

Buono, F. G., and Blackmond, D. G.\yjacs{2003}{125}{8978}
{8979}{Kinetic evidence for a tetrameric transition state in the 
asymmetric autocatalytic alkylation of pyrimidyl aldehydes}

Frank, F. C.\yjour{1953}{Biochim.\ Biophys.\ Acta}{11}{459}
{464}{On Spontaneous Asymmetric Synthesis}

Joyce, G. F., Visser, G. M., van Boeckel, C. A. A., van Boom, J. H.,
Orgel, L. E., and Westrenen, J.\ynat{1984}{310}{602}
{603}{Chiral selection in poly(C)-directed synthesis of oligo(G)}

Hegstrom, R. A.\yjour{1984}{Orig.\ Life}{14}{405}
{414}{Parity nonconservation and the origin of biological chirality --
theoretical calculations}

Kitamura, M., Suga, S., Oka, H., and Noyori, R.\yjacs{1998}{120}{9800}
{9809}{Quantitative analysis of the chiral amplification in the amino
alcohol-promoted asymmetric alkylation of aldehydes with dialkylzincs}

Kozlov, I. A., Pitsch, S., and Orgel, L. E.\ypnas{1998}{95}{13448}
{13452}{Oligomerization of activated D- and L-guanosine mononucleotides
on templates containing D- and L-deoxycytidylate residues}

Mathew, S. P., Iwamura, H., and Blackmond,
D. G.\yjour{2004}{Ang. Chem. Int. Ed.}{43}{3317}
{3331}{Amplification of enantiomeric excess in a proline-mediated reaction}

Mislow, K.\yjour{2003}{Collect. Czech. Chem. Commun.}{68}{849}
{864}{Absolute asymmetric synthesis: a commentary}

Nelson, K. E., Levy, M., and Miller, S. L.\yjour{2000}
{Proc.\ Nat.\ Acad.\ Sci.\ U.S.A.}{97}{3868}
{3871}{Peptide Nucleic Acids rather than RNA may have been
the first genetic molecule}

Plasson, R., Bersini, H., and Commeyras, A.\ypnas{2004}{101}{16733}
{16738}{Recycling Frank: spontaneous emergence of homochirality in
noncatalytic systems}

Pooga, M., Land, T., Bartfai, T., and
Langel, \"U.\yjour{2001}{Biomol.\ Eng.}{17}{183}
{192}{PNA oligomers as tools for specific modulation of gene expression}

Rasmussen, S., Chen, L., Nilsson, M., and
Abe, S.\yjour{2003}{Artif.\ Life}{9}{269}
{316}{Bridging nonliving and living matter}

Rikken, G. L. J. A. and Raupach, E.\ynat{2000}{405}{932}
{935}{Enantioselective magnetochiral photochemistry}

Rubenstein, E., Bonner, W. A., Noyes, H. P.,
and Brown, G. S.\ynat{1983}{306}{118}
{118}{Super-novae and life}

Sandars, P. G. H.\yoleb{2003}{33}{575}
{587}{A toy model for the generation of homochirality during polymerization}

Saito, Y. and Hyuga, H.\yjour{2004a}{J.\ Phys.\ Soc.\ Jap.}{73}{33}
{35}{Complete homochirality induced by the nonlinear autocatalysis
and recycling} (SH)

Saito, Y. and Hyuga, H.\yjour{2004b}{J.\ Phys.\ Soc.\ Jap.}{73}{1685}
{1688}{Homochirality proliferation in space}

Sato, I., Urabe, H., Ishiguro, S., Shibata, T., and
Soai, K.\yjour{2003}{Angew. Chem. Int. Ed.}{42}{315}
{317}{Amplification of chirality from extremely low to greater than
99.5\% {\it ee} by asymmetric autocatalysis}

Schmidt, J. G., Nielsen, P. E., \& Orgel, L. E.\yjacs{1997}{119}{1494}
{1495}{Enantiomeric cross-inhibition in the synthesis of oligonucleotides
on a nonchiral template}

Soai, K., Shibata, T., Morioka, H., and Choji, K.\ynat{1995}{378}{767}
{768}{Asymmetric autocatalysis and amplification of enantiomeric excess
of a chiral molecule}

Strong, W. M.\ynat{1898}{59}{53}
{54}{Stereochemistry and vitalism}

Thiemann, W.\yoleb{1984}{14}{421}
{426}{Speculations and facts on the possible inductions of chirality
through earth magnetic field}

Turner, P. C., McLennan, A. G., Bates, A. D., and
White, M. R. H.\ybook{2000}{Molecular biology}
{BIOS Scientific Publishers, Taylor \& Francis Group, London and New York}

Wattis, J. A. D. and Coveney, P. V.\poleb{2005}
{Symmetry-breaking in chiral polymerization}
Preprints available online at: \url{http://arXiv.org/abs/physics/0402091}

Woese, C.\ybook{1967}{The Genetic Code}{New York: Harper and Row}

\end{list}
\vfill\bigskip\noindent{\it
$ $Id: paper.tex,v 1.32 2005/05/14 01:10:58 brandenb Exp $ $}
\end{document}